\documentclass[prl,twocolumn]{revtex4}
\usepackage{epsfig}
\usepackage{bm}
\usepackage{amsmath}

\begin{document}

\title{Chaotic and turbulent dispersion of heavy inertial particles}

\author{A. Bershadskii}

\affiliation{
ICAR, P.O. Box 31155, Jerusalem 91000, Israel
}

\begin{abstract}
  Chaotic and turbulent dispersion of passive heavy inertial particles in homogeneous two-dimensional turbulence with Ekman drag has been studied using notions of the effective diffusivity and distributed chaos. Results of recent direct numerical simulations have been used for this purpose. 
  Effect of the Stokes number value on the applicability of these notions to the dispersion of passive heavy inertial particles has been briefly discussed.
\end{abstract}

\maketitle

  In order to introduce the notions of distributed chaos  and effective (chaotic/turbulent) diffusivity let us start 
from non-inertial passive scalar case. In this case the particle velocity ${\bf u}$ coincides with the fluid velocity ${\bf v}$. The later is described by the incompressible Navier-Stokes equations  \cite{my}
$$  
  \partial_t {\bf v} + ({\bf v}\cdot \nabla) {\bf v} = -\nabla p +\nu \nabla^2 {\bf v}+\mathbf{f}_{\bf v}, \eqno{(1)}
$$
$$  
  \nabla \cdot \bf{v}=0,   \eqno{(2)}
$$

\begin{figure} \vspace{-0.5cm}\centering
\epsfig{width=.45 \textwidth,file=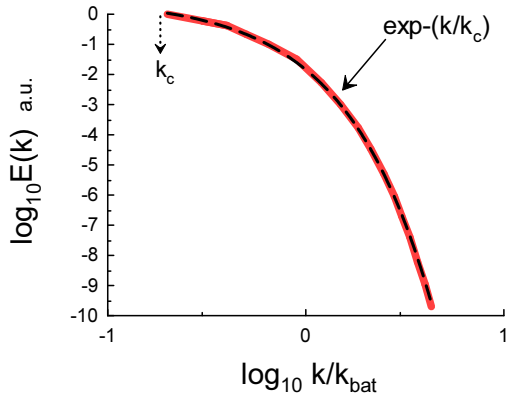} \vspace{-3.8cm}
\caption{Power spectrum of passive scalar fluctuations at $Re_{\lambda} = 8$ and $Sc = 1$.} 
\end{figure}
  
and equation
$$
\partial_t \theta + (\bf{v}\cdot\nabla)\theta= \kappa\nabla^2\theta+f_{\theta}  \eqno{(3)}
$$  
describes stirring of the passive scalar $\theta ({\bf x}, t)$ by the fluid velocity field ${\bf v} ({\bf x}, t)$. \\

   Let us recall the notion of effective (chaotic/turbulent) diffusivity $D$ (see, for instance, recent Ref. \cite{bof} and references therein).  A simplest estimation of the effective diffusivity is
$$
D=v_cl_c   \eqno{(4)}
$$
with $l_c$ as characteristic spatial scale and $v_c$ as characteristic velocity scale. In the wavenumber (the Fourier space) terms 
$l_c \propto 1/k_c$, then 
$$
D \propto v_ck_c^{-1}  \eqno{(5)}
$$
or, alternatively,
$$
v_c \propto D~k_c  \eqno{(6)}
$$

  It was recently shown by direct numerical simulations (DNS) \cite{kds} that at the onset of homogeneous isotropic turbulence (deterministic chaos) the spatial kinetic energy spectrum has exponential form 
$$
E(k) \propto \exp-(k/k_c)  \eqno{(7)}
$$    
  
   As one can see from figure 1 the same is true for corresponding spatial spectrum of the passive scalar fluctuations (the spectral data have been taken from Fig. 1a of the Ref. \cite{dsy} for small Taylor-Reynolds number $Re_{\lambda} =8$).  
The dashed curve in this figure indicates the exponential spectrum Eq. (7) ($k_{bat}$ is the Batchelor scale \cite{my} and the dotted arrow indicates position of the $k_c$ scale in the log-log scales).  \\

  For the so-called turbulent $Re_{\lambda}$ there is no single fixed parameter $k_c$. This parameter is fluctuating and in order to calculate the spectrum one should take an ensemble average
$$
E(k) \propto \int_0^{\infty} P(k_c) \exp -(k/k_c)dk_c  \propto \exp-(k/k_{\beta})^{\beta}  \eqno{(8)}
$$    
with certain distribution of the parameter $k_c$ - $P(k_c)$. One can also use a natural generalization of the original exponential spectrum to the stretched exponential one. Then asymptotical behaviour of the $P(k_c)$ at large $k_c$ follows from the Eq. (8) \cite{jon}
$$
P(k_c) \propto k_c^{-1 + \beta/[2(1-\beta)]}~\exp(-bk_c^{\beta/(1-\beta)}) \eqno{(9)}
$$
If the fluctuating $v_c$ has Gaussian distribution then for constant effective diffusivity one can conclude from the Eq. (6) that $k_c$ also has a Gaussian distribution.

   On the other hand, the distribution $P(k_c)$ given by Eq. (9) is Gaussian when $\beta = 2/3$. Hence
$$
E(k) \propto \exp-(k/k_{\beta})^{2/3}  \eqno{(10)}
$$
in this case.

   Figure 2 shows power spectrum for the passive scalar at $Re_{\lambda}=427$ (a three-dimensional fully developed steady homogeneous isotropic turbulence, the spectral data correspond to the DNS reported in the Ref. \cite{wg}).   The dashed curve in this figure indicates the stretched exponential spectrum Eq. (10). Since the scale $k_{\beta}$ separates the power-law ('-5/3') and the distributed chaos regions of scales it can be estimated using the equilibrium theory suggested in the Ref. \cite{b1}
$$
\ln(k_{bat}/k_{\beta}) \simeq 3  \eqno{(11)}
$$

\begin{figure} \vspace{-1.9cm}\centering
\epsfig{width=.45 \textwidth,file=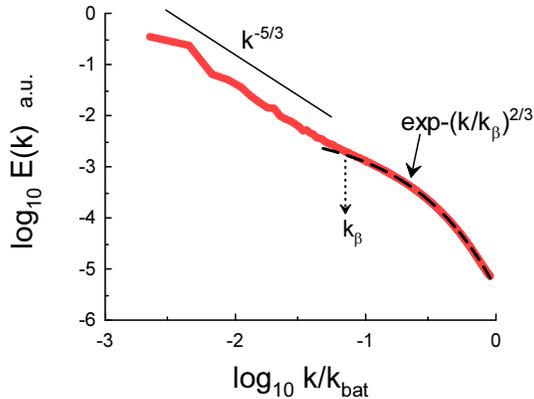} \vspace{-3.6cm}
\caption{Power spectrum of the passive scalar at $Re_{\lambda}=427$ and $Sc=1$.} 
\end{figure}
\begin{figure} \vspace{-0.5cm}\centering
\epsfig{width=.45\textwidth,file=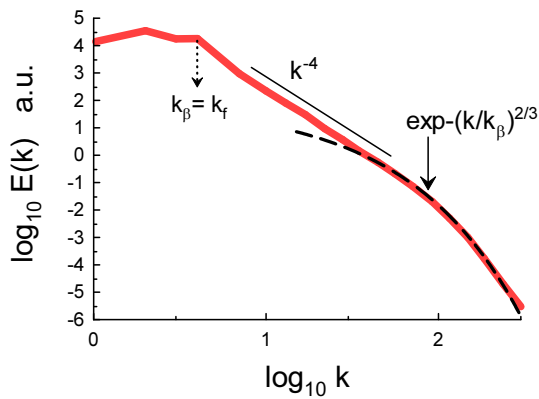} \vspace{-4.1cm}
\caption{Power spectrum of the of the particles velocity field ${\bf u} ({\bf x}, t)$ at $St = 3.4 \cdot 10^{-3}$ .} 
\end{figure}

  In the case of homogeneous isotropic three-dimensional turbulence the constant effective diffusivity can be supported by the Chkhetiani invariant  \cite{otto1}-\cite{b2}:
$$
I =     \int \langle {\bf v} ({\bf x},t) \cdot  {\boldsymbol \omega} ({\bf x} + {\bf r}, t) \rangle d{\bf r} \eqno{(12)}
$$
where ${\boldsymbol \omega}$ is vorticity and the brackets denote ensemble average. Namely
$$
D = C  \lvert I \rvert^{1/2}  \eqno{(13)}
$$  
here $C$ is a dimensionless constant.  For two-dimensional fluid motion the Chkhetiani invariant equals zero. However, an analogy of the Birkhoff-Saffman invariant 
$$
\mathcal{I} = \int r^2 \langle \omega ({\bf x},t) \cdot  \omega ({\bf x} + {\bf r}, t) \rangle d{\bf r} \eqno{(14)}
$$
\cite{saf}-\cite{d} can be used in the Eq. (13) instead of the $I$.\\

 In the case of the inertial particles, unlike the above discussed passive scalar mixing, the particles velocity field ${\bf u}$ does not coincides with the fluid velocity field ${\bf v}$ and the problem of dispersion of the inertial particles is more difficult. In order to simplify the problem some assumptions are usually used. First of all the particles can be also considered as passive and for a sufficiently dilute monodisperse suspension the inter-particle interactions can be ignored. The gravity can be also ignored if one is interested mainly in the turbulence effects. If the particles are small in comparison with the dissipative scale of the flow and heavy, i.e. their density $\rho_p \gg \rho_f$ (where $\rho_f$ is density of the fluid), then motion of a single particle can be described by equations
$$
\dot{{\bf x}} = {\bf u}, \eqno({15})
$$
$$
\dot{{\bf u}} = \frac{1}{\tau_p}\left[ {\bf v ({\bf x})} - {\bf u} \right] \eqno{(16)}
$$
where ${\bf x}$ is position of the particle and $\tau_p$ is so-called relaxation time. The inertial properties of the particles motion can be characterized by the Stokes number $St = \tau_p/\tau_{\eta}$, where $\tau_{\eta}= (\nu/\varepsilon)^{1/2}$ is the so-called Kolmogorov (or viscous) time scale. The inertia of the particles, relative to 
the fluid motion, is larger for larger Stokes numbers.  \\ 

   Even at these simplifications the problem of the particles dispersion by turbulent fluid motion is still a difficult one. Therefore a two-dimensional fluid turbulence is usually used for direct numerical simulations (see, for instance, reviews Refs. \cite{bof2},\cite{pan} and references therein). Keeping in mind possible applications for the oceanic and atmospheric flows the vorticity-streamfunction ($\omega$-$\psi$) form of the incompressible Navier-Stokes equations with the Ekman (linear) drag:
$$
\partial_t \omega + {\bf v} \cdot \nabla \omega = \nu \nabla^2 \omega -\alpha \omega + f_{\omega}  \eqno{(17)}
$$
$$
\nabla^2 \psi = \omega   \eqno{(18)}
$$
(where $f_{\omega}=-f_0 k_f \cos(k_f x)$ is the Kolmogorov 
forcing) was used for the direct numerical simulations reported in the recent Ref. \cite{mp}. 
\begin{figure} \vspace{-1.5cm}\centering
\epsfig{width=.45\textwidth,file=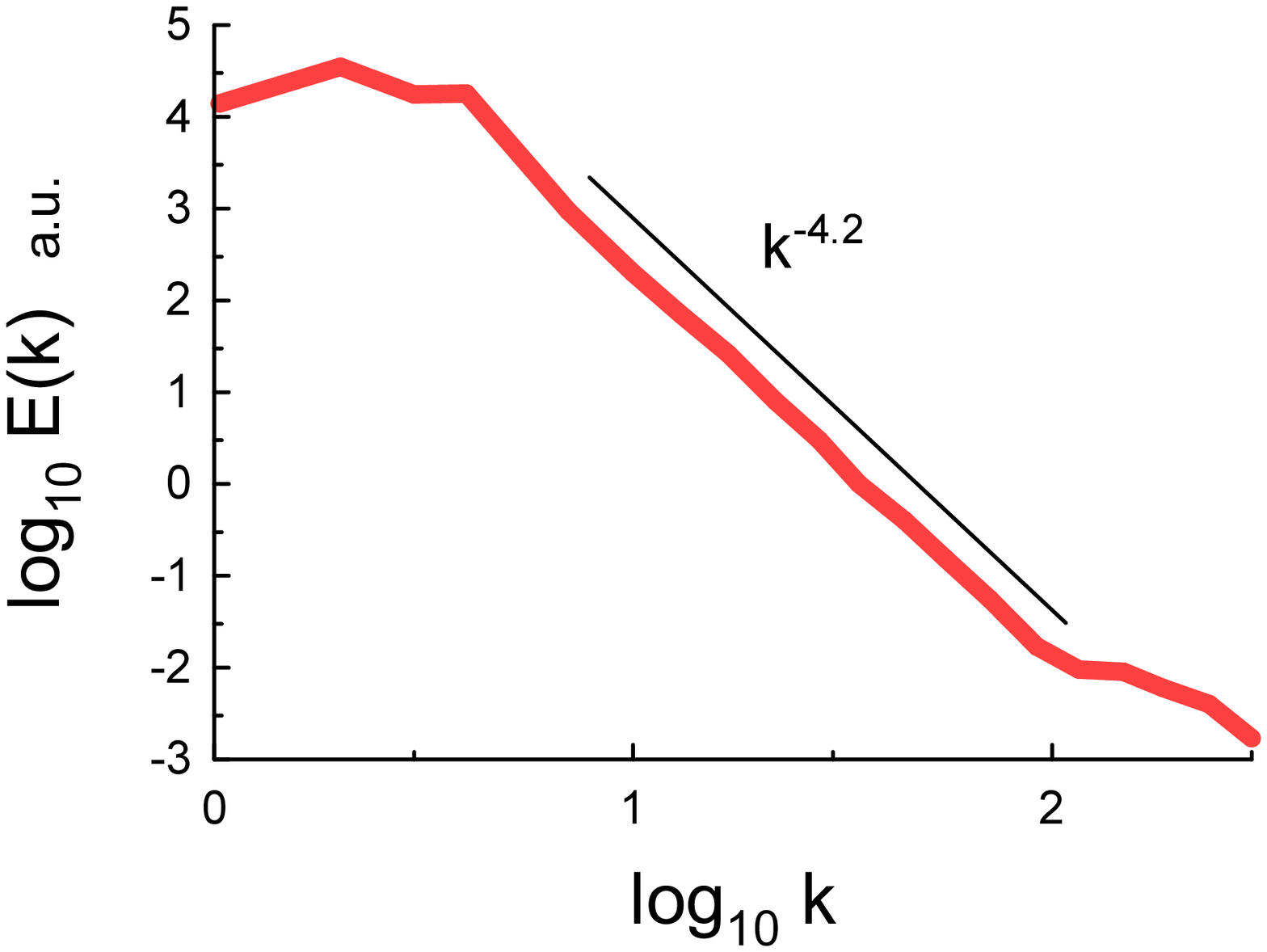} \vspace{-4cm}
\caption{As in the Fig. 3 but for $St= 1.7 \cdot 10^{-1}$. } 
\end{figure}
\begin{figure} \vspace{-0.3cm}\centering
\epsfig{width=.45\textwidth,file=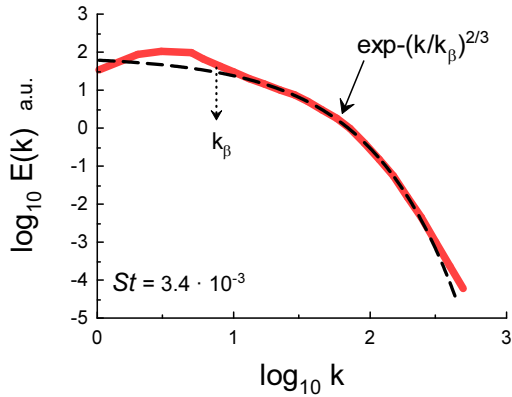} \vspace{-4.1cm}
\caption{Power spectrum of the of the particles density field $\rho ({\bf x}, t)$ at $St = 3.4 \cdot 10^{-3}$ .} 
\end{figure}
\begin{figure} \vspace{-1.7cm}\centering
\epsfig{width=.45\textwidth,file=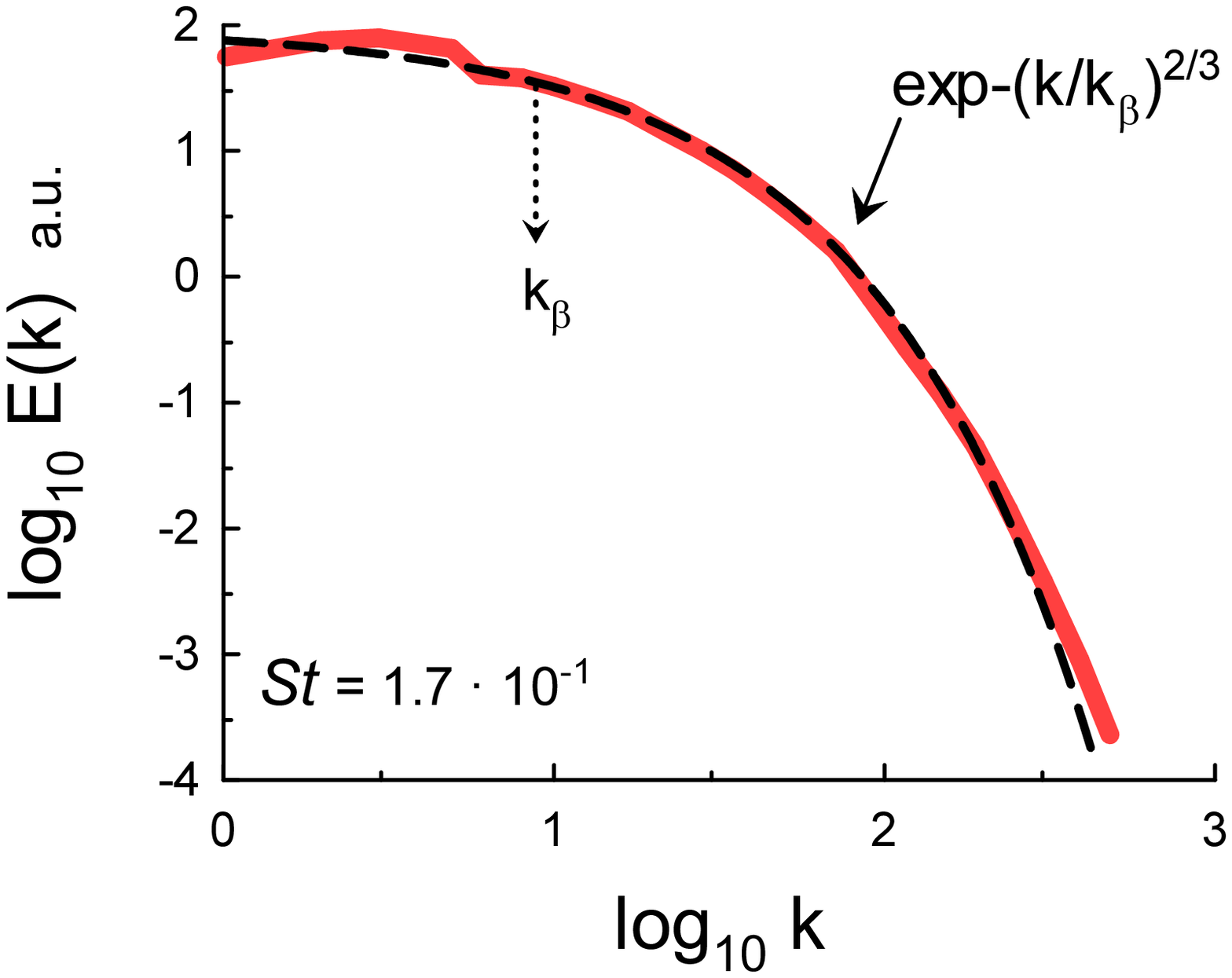} \vspace{-3.8cm}
\caption{As in the Fig. 5 but for $St= 1.7 \cdot 10^{-1}$. } 
\end{figure}
   The inertial particles density field $\rho$ was modelled in the Ref. \cite{mp} using the Eqs. (15),(16) and equations
$$
\partial_t \rho + \nabla \cdot (\rho {\bf u})  = 0  \eqno{(19)}
$$
$$
\partial_t (\rho {\bf u}) + \nabla \cdot (\rho {\bf u} \otimes {\bf u}) =  \frac{\rho}{\tau_p} \left[{\bf v}- {\bf u} \right]   \eqno{(20)}
$$
(where $\otimes$ denotes direct product of the vectors) with periodic boundary conditions.\\

  Figure 3 shows spatial power spectrum of the particles velocity field ${\bf u} ({\bf x}, t)$ obtained in the DNS \cite{mp} at Reynolds number $Re = 1319$, $\alpha = 0.01$ and $k_f =4$  (in the DNS spatio-temporal scales) and $St = 3.4 \cdot 10^{-3}$ . The spectral data were taken from Fig. 3a of the Ref. \cite{mp}. The dashed curve in this figure indicates the stretched exponential spectrum Eq. (10) and the dotted arrow indicates position of the $k_{\beta}$ scale.  This scale $k_{\beta} \simeq 4$, i.e. it is equal to the forcing scale $k_f$. Figure 6 shows analogous spectrum obtained at $St = 1.7 \cdot 10^{-1}$. In this case (for larger particles' inertia) the distributed chaos range is absent in the velocity field spectrum and the power-law spectrum $E \propto k^{-4.2}$ dominates the spectral decay.

    Figure 5 shows spatial power spectrum of the particles density field $\rho ({\bf x}, t)$ obtained in the same DNS at $St = 3.4 \cdot 10^{-3}$. The spectral data were taken from Fig. 3b of the Ref. \cite{mp}. The dashed curve in this figure indicates the stretched exponential spectrum Eq. (10) and the dotted arrow indicates position of the $k_{\beta}$ scale.  Figure 6 shows analogous spectrum obtained at $St = 1.7 \cdot 10^{-1}$. It is interesting that for the larger value of the Stokes number (i.e. for larger inertia of the particles) the distributed chaos range in the power spectrum of the particles density field is wider (cf. Figures 5 and 6). The extension of the distributed chaos range is in the direction of the small wavenumbers. For large wavenumbers applicability of the distributed chaos approach is restricted by the Kolmogorov (viscous) scale $\eta = (\nu^3/\varepsilon)^{1/4}$ for both values of the Stokes number.\\

I thank  T. Gotoh and T. Watanabe for sharing 
their data \cite{wg} and O.G. Chkhetiani and V. Yakhot for sending their papers.

\end{document}